# Bayesian analysis of extreme values in economic indexes and climate data: Simulation and application

Ali Reza Fotouhi

University of the Fraser Valley

33844 King Road, Abbotsford

BC, V2S 7M8

Canada

E-mail: ali.fotouhi@ufv.ca

**Abstract**

Mixed modeling of extreme values and random effects is relatively unexplored topic. Computational difficulties in using the maximum likelihood method for mixed models and the fact that maximum likelihood method uses available data and does not use the prior information motivate us to use Bayesian method. Our simulation studies indicate that random effects modeling produces more reliable estimates when heterogeneity is present. The application of the proposed model to the climate data and return values of some economic indexes reveals the same pattern as the simulation results and confirms the usefulness of mixed modeling of random effects and extremes. As the nature of climate and economic data are massive and there is always a possibility of missing a considerable part of data, saving the information included in past data is useful. Our simulation studies and applications show the benefit of Bayesian method to save the information from the past data into the posterior distributions of the parameters to be used as informative prior distributions to fit the future data. We show that informative prior distributions obtained from the past data help to estimate the return level in Block Maxima method and Value-at-Risk and Expected Shortfall in Peak Over Threshold method with less bias than using uninformative prior distributions

1. ## Introduction

Based on the reported news (https://www.abbynews.com/community/history-memories-of-1948-fraser-river-floods-still-run-strong/) Fraser valley in the province of British Columbia faced with a disaster when the Fraser River water level rises in 1948. Each year, when the snow melts quickly in the warm sun, the waters rise along the Fraser River, making their way to the Salish Sea. While this year was particularly difficult for many regions, the 1948 floods still hold the record as the region's worst flooding in living memory. From Chilliwack to Richmond, the deluge of water that rushed down the Fraser River in 1948 has yet to be surpassed. The statistics of that year were staggering, with more than 2,300 homes destroyed, 16,000 people forced to evacuate and many livestock and crops lost in a time when the area relied heavily on agriculture.

At its peak, the water level was measured at 7.6 meters. Many disasters like this can be addressed around the world. Analysis of extreme temperature and extreme precipitation may provide useful predictions to avoid huge damage to the nature and to human life.

Analysis of extreme values through statistical modeling uses extreme value theory and is an important topic in economic crises, climate forecasting, insurance, hydrology, and environmental applications where the risk of extreme events is of interest [1,2]. Manfred Gilli and Evis KÄellezi [3] focused on the use of extreme value theory to compute tail risk measures and the related confidence intervals, applying it to ES50, FTSE100, HS, Nikkei, SMI, and S&P500 stock market indexes. An insurance company may need to consider the potential risk of large claims cussed by earthquake and estimate the price of the insurance products to cover the potential insurance risk or calculation of risk in industry loss warranties [4]. Researchers may decide to predict the probability of the next flood event in a location to reduce the loss caused by future flood [5,6]. An engineer may wish to estimate the probability of an extreme wind speed in a location that can damage environment [7,8].

Although descriptive and exploratory analysis of past data can help to predict the situation in the near future but statistical modeling based on distributional theory can give a prediction with some measure of reliability. There are always unobservable variables that may affect the extremes within blocks or some periods of time. Statistical modeling of the extreme values allows incorporating these unobservable factors through a mixed model of random effects and extremes to produce more reliable predictions. Up to our knowledge, very little work has been done on random effects modeling in extreme value data analysis.

There are many problems for which there do not exist mathematical solution. Simulation can be used to compare different approaches or different statistical models when a mathematical proof is not available. In simulation studies we know the true mode, whereas in empirical analysis we do not. We can never be certain if the simulation results are relevant in practice but if empirical analysis reveals the same patterns then we can be reasonably confident, i.e. we can rely on the pragmatic considerations in evaluating the importance of simulation results.

As the nature of economic indexes and climate data is dynamic and the data are observed over time for a long period of time, there is always a risk of missing all or part of the past data. If a small fraction of past data is lost we may employ some statistical methods to estimate the missing data. If a large amount of past data is lost or in a very bad scenario all of the past data are lost then prediction is not reliable or is completely impossible. One way to be on the safe side is to save the information about the past data and use them together with the current data for prediction. Maximum likelihood method estimates the parameters and provides predictions using available data. If the past data are lost then this method do not use the information included in the past data. Bayesian analysis of the data enables us to save the information included in the past data in posterior distributions of the parameters and, in case of losing the past data, use them as prior distributions for the parameters to fit current data. In this case, if not all, but part of the information included in the past data can be used for the analysis of the current data.

This article will be continued as following. In section 2 we discuss the theory including distributions and approaches. In section 3 we introduce the idea of random effects modeling. In section 4 we discuss estimation and computational methods. In section 5 we report some simulation results. In section 6 we report the application of the proposed models in analyzing return values of six economy indexes, maximum temperature, and maximum precipitation. In section 7 we discuss the use of informative prior distributions. In section 8 we report the summery of the analysis.

## 2  Theory

The field of extreme value theory was initiated by Leonard Tippett lived from 1902 to 1985. This theory provides the probability models needed for statistical modeling of the extreme values [1], [9,10]. There are generally two ways of identifying extreme values in a dataset. The Block Maxima (BM) method involves splitting the dataset into blocks of fixed size, and finding the maximum or minimum value in each block. For example, consider a set of financial series of daily returns. We divide the daily returns in $n$ blocks of equal length, say years, and collect the maximum (or minimum) of return values in each year. In order to estimate useful characteristics of extremes such as quantiles we need to know the distribution of the extremes. We will discuss the distribution of maximums; as for minimums the distribution of negative of minimums is the same as the distribution of maximums. The distribution of the original observations $x_{i1}, x_{i2}, \ldots, x_{in}$ is unknown but according to Fisher and Tippett [11] and Gnedrenko [12] the limiting distribution of the maximums $y_i = max(x_{i1}, x_{i2}, \ldots, x_{in})$ in blocks, if it exists as the block size approaches infinity, belongs to the Generalized Extreme Value (GEV) family of distributions. This theorem does not guaranty that the limiting distribution of maximums always exists. Frechet, Weibull, and Gumbel are special cases of GEV distribution. The cumulative distribution of GEV is of the form

$$H(y_i|\xi,\mu,\sigma) = \begin{cases} e^{-\left[1+\xi\left(\frac{y_i-\mu}{\sigma}\right)\right]^{\frac{-1}{\xi}}} & \xi \neq 0 \\ e^{-e^{-\left(\frac{y_i-\mu}{\sigma}\right)}} & \xi = 0 \end{cases} \quad (1)$$

where $-\infty < y_i < \mu - \frac{\sigma}{\xi}$ if $\xi < 0$, $-\infty < y_i < +\infty$ if $\xi = 0$, $\mu - \frac{\sigma}{\xi} < y_i < +\infty$ if $\xi > 0$ and $-\infty < \mu < +\infty, \sigma > 0$. The expected value of $y_i$ is

$$E(y_i) = \begin{cases} \mu + \sigma(\Gamma(1-\xi)-1)/\xi & if\ \xi \neq 0, \xi < 1 \\ \mu + \sigma\gamma & if\ \xi = 0 \\ \infty & if\ \xi \geq 1 \end{cases}$$

where $\gamma$ is Euler's constant.

The parameters $\mu$ and $\sigma$ are the location and scale parameters that normalize the data and $\xi$ is the shape parameter of the GEV distribution. A negative value of $\xi$ implies a Weibull distribution, a positive value implies a Frechet distribution, and a zero value implies a Gumbel distribution. Block size is an important factor in using this family of distributions for analysis of extremes.

The second method of finding extreme values is using the Peak-Over Threshold (POT) method. In this method we set a fixed threshold $u$ and find all values above this value. Similar to BM methods the distribution of the original observations $x_1, x_2, x_3, \ldots$ is unknown but according to Pickands [13] and Balkema and de Hann [14], for a reasonably large threshold $u$, the distribution of $y_1, y_2, y_3, \ldots$ where $y_i = x_i - u$ is well approximated by the Generalized Pareto Distribution (GPD). The cumulative distribution of GPD family, in general, is

$$F(y_i|\xi,\mu,\sigma) = \begin{cases} 1 - \left[1 + \xi\left(\frac{y_i-\mu}{\sigma}\right)\right]^{-\frac{1}{\xi}} & \xi \neq 0 \\ 1 - e^{-\left(\frac{y_i-\mu}{\sigma}\right)} & \xi = 0 \end{cases} \qquad (2)$$

where $-\infty < y_i < +\infty$ if $\xi = 0$, $y_i > \mu - \frac{\sigma}{\xi}$ if $\xi > 0$, $y_i < \mu - \frac{\sigma}{\xi}$ if $\xi < 0$ and

$-\infty < \mu < +\infty, -\infty < \xi < +\infty, \sigma > 0, E(y_i) = \mu + \sigma/(1-\xi)\ if\ \xi < 1$.

The parameters μ and σ are the location and scale parameters that normalize the data and $\xi$ is the shape parameter of GPD. If both $\mu$ and $\xi$ are zero GPD is equivalent to exponential distribution and if $\mu = \frac{\sigma}{\xi}$ and $\xi > 0$ then GPD is equivalent to Pareto distribution. To use GPD for modeling extremes we set $\mu = u$ and use the observations exceed the threshold $u$. In this case, the location parameter or equivalently the threshold is fixed and only the scale and shape parameters are estimable. We should select a large threshold such that the asymptotic approximation holds [13], [14]. On the other hand, the threshold should not be assumed very large otherwise only few observations would exceed the threshold. Other than some graphical method [3] there is no proved method for identifying the threshold and this makes the use of the POT method more challenging than using BM method. However, depending on the purpose of the analysis we should select either of two methods.

There are advantages and disadvantages in using BM and POT methods. The advantage of using BM method is that it probably removes the dependences within block. The disadvantages are excluding some large values such as second highest in blocks and the selection of the block size. If the block size is too large we lose some data and if the block size is too small we cannot use the GEV distribution as the limiting distribution. The advantage of POT method is avoiding elimination of large observations that are close to the maximum in a neighborhood. The disadvantage is the difficulty in the selection of the threshold $u$. Selection of any of these methodologies depends on the data and the interpretation that we are expecting from the analysis.

One of the most frequent questions in the analysis of extreme values is the level that is expected to be exceeded in one out of $k$ periods of equal lengths. In economy return level can be interpreted as the maximum loss in analysis of indexes. In such analysis the parameters of interest for GEV distribution is $R^k = H^{-1}\left(1 - \frac{1}{k}\right)$. It can be shown that

$$R^k = \begin{cases} \mu - \frac{\sigma}{\xi}\left[1 - \left(-\log\left(1 - \frac{1}{k}\right)\right)^{-\xi}\right] & ;\ \xi \neq 0 \\ \mu - \sigma \log\left(-\log\left(1 - \frac{1}{k}\right)\right) & ;\ \xi = 0 \end{cases} \qquad (3)$$

A value of $R^k$ of $E$, in the analysis of maximum values of $y$ over the equal period of time $T$, means that the maximum loss observed during a period of time $T$ exceeds $E$. When estimation of return level is of interest BM method is used.

Another important measure in the analysis of extreme values, mainly in finance, is a value that the random observation exceeds that value with a low probability such as 1% or 5%. Two of those measures are Value At Risk (*VAR*) and Expected Shortfall (*ES*). For a random variable x and for probability $p$, VAR is defined as $VAR_p = G^{-1}(1-p)$ where $G$ is the cumulative distribution function of the random variable $x$. Following Gilli and Kaellezi [3] we can show that

$$VAR_p = u + \frac{\sigma}{\xi}\left(\left(\frac{n}{N_u}p\right)^{-\xi} - 1\right) \qquad (4)$$

Where $n$ is the total number of observations $x_1, x_2, x_3, \ldots$ and $N_u$ is the number of observations $y_1, y_2, y_3, \ldots$ exceed the threshold $u$. ES is defined as $ES_p = E(X|X > VAR_p)$ and can be shown that

$$ES_p = (VAR_p + \sigma - \xi u)/(1 - \xi) \qquad (5)$$

The interpretation of $VAR_p$ and $ES_p$ in the analysis of daily return value of an index is that, with probability $p$, the tomorrow's loss will exceed the value $VAR_p$ and the related expected loss, the average loss where the losses exceed $VAR_p$, is $ES_p$. POT approach is used when estimation of quantiles or expected shortfall is of interest.

### 3    Random effects model

There are always unobservable factors that may affect the extremes located in one cycle differently from another cycle. Unobservable factors may also affect the return value of on economic index differently from another index. This phenomenon produces heterogeneity in data and needs to be considered in statistical modeling of extremes. Introducing random effects in the parameter of GEV distribution that vary between indexes or cycles control the heterogeneity exists in data. We assume a specific distribution, usually normal distribution, for the random effects and mix it with GEV distribution. For details about mixed effects models see [15]. Random effects modeling is widely used in statistical modeling bur very little work has been done on random effects modeling in the analysis of extreme values.

Suppose there exist $I$ periods that may include one or more blocks and $y_{il}$ represents the $l^{th}$ extreme in the $i^{th}$ period where $l = 1,2,\ldots,n$ and $i=1,2,\ldots,I$. To control the heterogeneity between periods we propose to add a random component to the location parameter and consider $\mu_i = \mu + \delta_i$ as the location parameter for the extremes in the $i^{th}$ period. We assume $\delta_i$ has a normal distribution with mean zero and variance $\tau^2$. If $\delta_i = 0$ for all $i$, then we call the model a fixed effects model. As the mean of GEV distribution and $R^k$ are linear functions of $\mu$, the random effects component is actually added linearly to the mean and $R^k$. If $\tau^2$ is estimated significantly positive it indicates that heterogeneity exists and is captured by the mixed model. In this setting we actually assume that the location of extreme values is a random variable with mean $\mu$ and variance $\tau^2$. It is reasonable and frequently used in applications that conditional on random effects $\delta_i$ response variables are independent and therefore the joint probability density function (pdf) of $y_i = (y_{i1}, y_{i2}, \ldots, y_{i,n})$ is given by

$$L_i(\xi, \mu_i, \sigma) = \prod_{l=1}^{n} f(y_{il}|\xi, \mu_i, \sigma, \delta_i) \qquad (6)$$

where for GEV presented in equation (1) we have

$$f(y_{il}|\xi, \mu, \sigma, \delta_i) = \begin{cases} exp(-(1+\xi z_{il})^{-1/\xi})(1+\xi z_{il})^{-1-\frac{1}{\xi}}, & \xi \neq 0 \\ exp(-z_{il} - exp(-z_{il})), & \xi = 0 \end{cases}$$

$$\text{where } z_{il} = \frac{(y_{il}-\mu_i)}{\sigma} \ ; \ \mu_i = \mu + \delta_i \ ; \ \delta_i \text{ are } i.i.d \ N(0,\tau^2) \tag{7}$$

The expected value of the likelihood in equation (6) over the distribution of the random effects $\delta_i$ is given by

$$E\big(L_i(\xi,\mu,\sigma,\tau)\big) = \int_{-\infty}^{+\infty} L_i(\xi,\mu_i,\sigma)g(\delta_i)d(\delta_i)$$

Where $g(\delta_i)$ is the *pdf* of a normal distribution with mean zero and variance $\tau^2$. The total likelihood function over *I* periods is given by

$$L(\xi,\mu_i,\sigma,\tau) = \prod_{i=1}^{I} E\big(L_i(\xi,\mu_i,\sigma,\tau)\big) \tag{8}$$

Taking logarithm from equation (8) and substituting from (5) the log-likelihood function is given by

$$l(\xi,\mu,\sigma,\tau) = \sum_{i=1}^{I} log\big(\int_{-\infty}^{+\infty} \prod_{l=1}^{n_i} f(y_{il}|\xi,\mu_i,\sigma,\delta_i) \ g(\delta_i)d(\delta_i)\big). \tag{9}$$

It is probable that heterogeneity affects both the location and scale parameters as $\mu_i = \mu + \delta_{i1}$ and $\sigma_i = \sigma + \delta_{i2}$. In this case, the joint *pdf* of $\delta_{i1}$ and $\delta_{i2}$ should be considered instead of $g(\delta_i)$ and we have double integration in equation (9).

Caution should be done in the interpretation of the parameters in a random effects model. In fixed effects model, parameters have population average interpretations. In random effects modeling we have $\mu_i = \mu + \delta_i$ and therefore $E(\mu_i) = \mu$. Therefore the estimate of $\mu$ in random effects modeling is interpreted as the mean of location parameters over all periods. Random effects model is useful when period or subject specific effect on the estimation of the parameter is of interest.

### 4 Estimation and computational methods

In a classical statistical modeling the parameters $\xi, \mu, \sigma,$ and $\tau$ are estimated by maximizing $l(\xi,\mu,\sigma,\tau)$ in equation (9). However, equation (9) does not have a closed form and numerical integration should be used for estimation. The computational difficulties in using the Maximum Likelihood Estimation (MLE) method for mixed models and the need of saving the information included in past data motivate us applying Bayesian method using Markov chain Monte Carlo (MCMC) procedure. The *MCMC procedure* from SAS software begins with prior information about the parameters $\theta = (\xi,\mu,\sigma,\tau)$ present in the model through the prior density $p(\theta)$. It then uses the sampling density $f(y|\theta)$, Bayes' theorem, and Metropolis algorithm [16,17,18] to update the prior information $p(\theta)$ to a posterior density function of the parameters as

$$p(\theta|y) = \frac{f(y|\theta)p(\theta)}{\int f(y|\theta)p(\theta)d\theta} . \tag{10}$$

The point estimates of the posterior mean and the Highest Posterior Density (HPD) interval, which is the one with the smallest interval width among all credible intervals (*CI*) of the parameters, are calculated in *MCMC procedure*. A credible interval *(a,b)* is defined as

$$p(\theta \in CI|y) = \int_a^b p(\theta|y)d\theta.$$

Deviance Information Criterion (*DIC*) is a model assessment criterion in Bayesian framework. It is an alternative criterion to Akaike's and Bayesian information criteria (*AIC* and *BIC*). *DIC* has the flexibility that can be applied to models that are not nested or used for data that are not identically independently distributed. Unlike *AIC* and *BIC* the criterion *DIC* does not require maximization of the likelihood function but cannot be used for testing the nested models. A smaller value of *DIC* indicates a better fit. If the difference between *DIC* of two models is less than 5 there is no serious difference between two models, between 5 and 10 model with smaller *DIC* is preferred but the difference is not serious, and greater than 10 indicates that model with smaller *DIC* is the preferred model.

To use the *MCMC procedure* we need prior distributions of the parameters. Jeffreys' prior [19] and flat prior [20] have been considered as uninformative priors. If the analyst has prior information about the parameter and use it to establish a prior density function the resulting prior is called informative prior. We consider the following uninformative flat priors for the model presented in equation (9).

$$\xi \sim uniform(-10000, 10000) \qquad (11)$$

$$\mu \sim uniform(-10000, 10000)$$

$$\sigma \sim gamma(shape = 0.0001, scale = 10000)$$

$$\tau^2 \sim igamma(shape = 0.0001, scale = 0.0001)$$

For future simulations and applications we use *MCMC procedure* and assume uninformative prior distributions for the parameters $\theta = (\xi, \mu, \sigma, \tau)$ presented in (11). We produce 3000 pre-samples from posterior distributions of the parameters to eliminate the effects of the initial values given to the parameters and the assumed seed to start the Markov chain. This step actually set up the Markov chain for producing the main samples. Then we produce 20000 samples from the posterior distributions and take the average of every fifth sampled values to reduce the correlation between the successive sampled values. The estimate of each parameter will be the average of 4000 samples obtained from the posterior distribution. These specifications will be used in all future computations.

## 5  Simulation study

In this section we present a simulation study for two purposes. Firstly, we evaluate if GEV is an adequate family of distributions for analyzing yearly maximums. Although we mentioned that Fisher and Tippett [11] and Gnedrenko [12] proved that the limiting distribution of the maximums in blocks belongs to the GEV family of distributions but we need to check this through a simulation study to be sure that GEV distribution is an adequate distribution for the structure of the data we want to analyze. Secondly, we want to investigate the importance of the random effects modeling in the analysis of extremes when GEV distribution is used. This simulation study helps to check the methodology that we will use in application. The first simulation is set up as following. We produce random number, analogous to daily return value of an economic index, for 10 periods each includes 5 years. We introduce a normal random variable that varies between periods to define the heterogeneity of observations among periods. This allows us to evaluate the performance of GEV distribution in the situation that unobservable factors affect the maximums in one period differently from another period. The data are simulated according to the following steps.

> i) *set a fixed seed for random number generation*
> ii) *set $i = 1$*
> iii) *generate $e_i \sim N(mean = 0, sd = 1)$*
> iv) *generate $x_{il} \sim N(mean = 0.02 + e_i, sd = 1.24)$ for $l = 1,2, \dots ,18000$*
> v) *if i=10 stop, otherwise add 1 to i and repeat steps (ii) and (iii)*

The values 0.02 and 1.24 in step (*iv*) are taken from the combination of historical daily return values of six indexes EuroXX, FTS, HS, Nikkei, SMI, and SP from 1960 to 2004. We collect the maximum of the generated random numbers in each year. The time series plot of the 50 simulated maximums is shown in Figure 5.1. We fit the GEV distribution to the simulated maximums with and without random effects. The quantile plot reported in Figure 5.2 indicates that year is an adequate block size and the GEV model fits the yearly maximum values well. As it is possible that the heterogeneity affects both the location and scale parameters we fit two random effects models. The first random effects model assumes a random components added to the location parameter as $\mu_i = \mu + \delta_i$ where $\delta_i \sim N(0, \tau^2)$. In the second random effects model we include correlated random effects components in the location and scale parameters as $\mu_i = \mu + \delta_{1i}$ and $\sigma_i = \sigma + \delta_{2i}$. We consider a bivariate normal distribution for the random vector $(\delta_{1i}, \delta_{2i})$ as following

$$(\delta_{1i}, \delta_{2i}) \sim MVN(\theta, \Sigma) \; ; \; \theta = (\theta_1, \theta_2) \; ; \; \Sigma = \begin{bmatrix} \tau_1^2 & \rho\tau_1\tau_2 \\ \rho\tau_1\tau_2 & \tau_2^2 \end{bmatrix}$$

This setting allows modeling the possible correlation, $\rho$, between the two random components $\delta_{1i}$ and $\delta_{2i}$. We assure that $\sigma_i$ is positive during the estimation process. The results from fitting three models are shown in Table 5.1. The variances of the random effects are estimated positive with 95% probability in both random effects models. The random effects models produce different parameter estimates from fixed effects model. The difference is more evident in the estimation of $R^{10}$. The empirical return rate for the 50 years simulated data is 3.81 with a 95% confidence interval (3.50,4.12). The estimate of $R^{10}$ by location random effects model is much closer to its empirical value than the estimate from fixed effects model. Actually 3.81 is not located in the *95% HPD* interval (4.83,6.17) obtained from the fixed effects model while it is included in the *95% HPD* interval (3.53, 4.96) obtained from the location random effects model. *DIC* and *log-likelihood* for location random effects model are 61.23 and -25.62 while for fixed effects model are 147.58 and -75.00, respectively. Both *DIC* and *log-likelihood* statistics confirm that the location random effects model fit the simulated maximum better than the fixed effects model. The estimation of $R^{10}$ is also closer to its empirical estimate when random effects present in both location and scale parameters as compare to the fixed effects model. *DIC* and *log-likelihood* for location and scale random effects model are 53.90 and -21.61, respectively. This indicates that the location and scale random effects model fits the data better than the location random effects model. The correlation, $\rho$, between the random components $\delta_{1i}$ and $\delta_{2i}$ is 0.20, which indicates a low positive correlation. This simulation supports applying the random effects and fixed effects GEV models to the yearly maximums collected from observations generated by a normal distribution within heterogeneous periods but there are some evidences that random effects models estimate the return level more accurately.

**Figure 5.1:** The time series plot of the 50 simulated maximums.

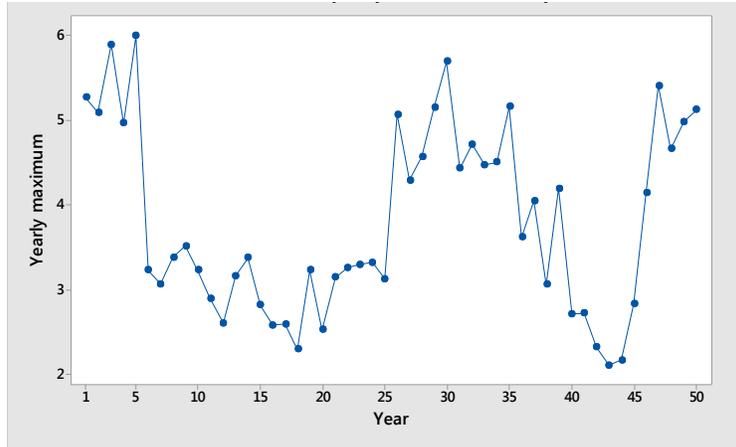

**Figure 5.2:** Quantile plot of the simulated maximums.

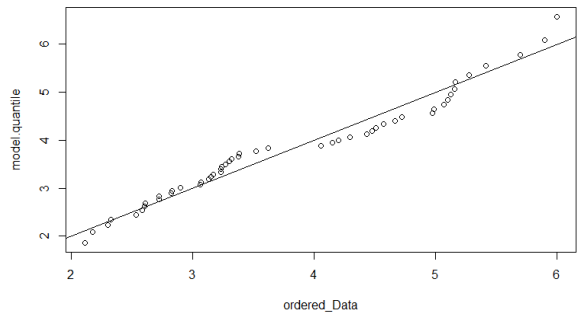

**Table 5.1:** Result from applying the block maxima method to simulated data.

| Random effects | Parameter | Mean | SD | LB* | UB* |
|---|---|---|---|---|---|
| None | $\xi$ | -0.10 | 0.18 | -0.45 | 0.24 |
| | $\mu$ | 3.33 | 0.17 | 3.00 | 3.65 |
| | $R^{10}$ | 5.39 | 0.37 | 4.83 | 6.17 |
| | $\sigma$ | 1.00 | 0.14 | 0.75 | 1.28 |
| In location | $\xi$ | -0.17 | 0.21 | -0.60 | 0.23 |
| | $\mu$ | 3.55 | 0.34 | 2.85 | 4.21 |
| | $R^{10}$ | 4.29 | 0.36 | 3.53 | 4.96 |
| | $\sigma$ | 0.39 | 0.07 | 0.26 | 0.52 |
| | $\tau^2$ | 1.38 | 0.86 | 0.37 | 2.89 |
| In location and scale | $\xi$ | -0.25 | 0.19 | -0.60 | 0.00 |
| | $\mu$ | 5.04 | 0.31 | 4.50 | 5.72 |
| | $R^{10}$ | 4.66 | 0.33 | 3.98 | 5.28 |
| | $\sigma$ | 0.05 | 0.08 | 0.00 | 0.23 |
| | $\theta_1$ | -1.40 | 0.48 | -2.30 | -0.40 |
| | $\theta_2$ | 0.42 | 0.18 | 0.05 | 0.76 |
| | $\tau_1^2$ | 1.27 | 0.79 | 0.33 | 2.67 |
| | $\rho\tau_1\tau_2$ | 0.10 | 0.22 | -0.34 | 0.50 |
| | $\tau_2^2$ | 0.20 | 0.13 | 0.05 | 0.42 |

*LB and LU are Lower and upper 95% HPD interval.

The second simulation has been set up for the elimination of the possible sampling error. We have replicated the previous simulation for 25 times and calculated the mean, standard deviation, and 95% confidence interval for the parameters over the replications. Each of the 25 simulated samples is generated with and without random effects and for each case the data are fitted with and without random effects models. The results are reported in Table 5.2. For the case that data are generated without random effects, both location random effects and fixed effects models estimate the parameters almost identical. The *log-likelihood* and *DIC* statistics are almost equal for both models. This indicates that even when the data are homogeneous among periods the random effects GEV model is safe to be used for modeling yearly maximums. For the case that data are generated with random effects, the location random effects model estimates the parameters differently from the fixed effects model. The variance of the location random effects is estimated $\tau^2 = 1.15$ with 95% confidence interval (0.85,1.46) indicating that the heterogeneity among the periods is well captured by the proposed model. The *log-likelihood* for the location random effects model is -31.31 and for the fixed effects model is -69.00. This indicates a better fit with location random effects model. The *DIC* criterion for the location random effects model is 72.70 while for the fixed effects model is 140.65. This criterion strongly prefers the location random effects model to the fixed effects model. Similar comparison indicates that the location and scale random effects model performs better than the fixed effects model. The average of empirical $R^{10}$ over the 25 simulated samples is calculated as 3.70 with a 95% confidence interval (3.64, 3.77). The estimate of $R^{10}$ produced by location random effects model is 4.37 and by location and scale random effects model is 4.64 while by fixed effects model is 5.11. Therefore, both random effects models produced better estimates for $R^{10}$ than the fixed effects model. The empirical $R^{10}$ for each of 10 periods are reported in Table 5.3 for the interpretation within periods individually.

**Table 5.2:** Simulation result for 25 simulated random samples. Figures shown are the mean of estimates over the 25 replications.

| Data | Random effects | Parameter | Mean | SD | LB* | UB* |
|---|---|---|---|---|---|---|
| Generated without random effects | None | $\xi$ | -0.10 | 0.10 | -0.15 | -0.05 |
| | | $\mu$ | 3.47 | 0.05 | 3.44 | 3.50 |
| | | $R^{10}$ | 4.35 | 0.15 | 4.29 | 4.40 |
| | | $\sigma$ | 0.43 | 0.05 | 0.41 | 0.45 |
| | | $ll^{**}$ | -31.81 | 5.65 | -34.15 | -29.47 |
| | | DIC | 66.33 | 11.35 | 61.64 | 71.02 |
| | In location | $\xi$ | -0.09 | 0.15 | -0.15 | -0.03 |
| | | $\mu$ | 3.47 | 0.05 | 3.44 | 3.50 |
| | | $R^{10}$ | 4.33 | 0.15 | 4.28 | 4.39 |
| | | $\sigma$ | 0.42 | 0.05 | 0.40 | 0.44 |
| | | $\tau^2$ | 0.03 | 0.00 | 0.02 | 0.04 |
| | | $ll$ | -30.58 | 5.35 | -32.80 | -28.37 |
| | | DIC | 67.03 | 10.65 | 62.64 | 71.42 |
| Generated with random effects | None | $\xi$ | -0.19 | 0.15 | -0.26 | -0.12 |
| | | $\mu$ | 3.31 | 0.35 | 3.16 | 3.46 |
| | | $R^{10}$ | 5.11 | 0.45 | 4.93 | 5.30 |
| | | $\sigma$ | 0.98 | 0.25 | 0.87 | 1.09 |
| | | $ll$ | -69.00 | 12.7 | -74.25 | -63.76 |
| | | DIC | 140.65 | 25.4 | 130.163 | 151.141 |
| | In location | $\xi$ | -0.10 | 0.15 | -0.16 | -0.03 |
| | | $\mu$ | 3.49 | 0.35 | 3.35 | 3.64 |
| | | $R^{10}$ | 4.37 | 0.35 | 4.23 | 4.52 |
| | | $\sigma$ | 0.43 | 0.05 | 0.40 | 0.45 |
| | | $\tau^2$ | 1.15 | 0.75 | 0.85 | 1.46 |
| | | $ll$ | -31.31 | 6.60 | -34.04 | -28.58 |
| | | DIC | 72.70 | 13.3 | 67.21 | 78.19 |
| | In location and scale | $\xi$ | -0.21 | 0.22 | -0.30 | -0.12 |
| | | $\mu$ | 3.48 | 3.79 | 1.92 | 5.05 |
| | | $R^{10}$ | 4.64 | 2.80 | 3.48 | 5.80 |
| | | $\sigma$ | 0.43 | 1.38 | -0.14 | 1.00 |
| | | $\theta_1$ | 0.02 | 3.75 | -1.53 | 1.57 |
| | | $\theta_2$ | 0.12 | 1.37 | -0.45 | 0.69 |
| | | $\tau_1^2$ | 1.10 | 0.64 | 0.83 | 1.36 |
| | | $\rho\tau_1\tau_2$ | 0.01 | 0.06 | -0.02 | 0.03 |
| | | $\tau_2^2$ | 0.20 | 0.02 | 0.19 | 0.20 |
| | | $ll$ | -32.85 | 6.77 | -35.64 | -30.05 |
| | | DIC | 75.98 | 14.43 | 70.03 | 81.94 |

*LB and LU are Lower and upper 95% confidence interval. **log-likelihood.

**Table 5.3:** Empirical return rate $R^{10}$ calculated within periods over 25 replications.

| Cycle | Mean | SD | LB* | UB* |
|---|---|---|---|---|
| 1 | 3.59 | 1.12 | 3.39 | 3.79 |
| 2 | 3.56 | 1.08 | 3.37 | 3.75 |
| 3 | 3.68 | 0.95 | 3.52 | 3.85 |
| 4 | 3.75 | 0.92 | 3.59 | 3.91 |
| 5 | 3.39 | 1.03 | 3.21 | 3.57 |
| 6 | 3.67 | 1.04 | 3.48 | 3.85 |
| 7 | 3.78 | 1.08 | 3.59 | 3.97 |
| 8 | 3.73 | 1.13 | 3.53 | 3.93 |
| 9 | 3.73 | 1.03 | 3.55 | 3.91 |
| 10 | 4.10 | 1.20 | 3.89 | 4.31 |

*LB and LU are Lower and upper 95% confidence interval.

In the third simulation we produce different data sets with different value of the standard deviation of the random effects ($\tau = 0,1,2,3, and\ 4$ ) to induce different levels of heterogeneity through the location parameter. We fit the GEV model to each simulated data set with and without random effects. Table 5.4 reports 95% confidence intervals for the empirical and estimated values of $R^{10}$. This table shows that when data are produced with no random effects ($\tau = 0$) both random effects and fixed effects models estimate $R^{10}$ almost identical. As the standard deviation of the random effects increases the difference between the estimates from the random effects model and fixed effects model increases. This is while the estimates from the random effects model are consistently close to the empirical estimates and the estimate from the fixed effects model gets larger than the empirical estimates. This simulation suggests using the random effects model in estimation of the return level even when the data are homogeneous. The importance of using random effects becomes more serious when the level of heterogeneity gets larger.

**Table 5.4:** Simulation result for different level of heterogeneity.

| $\tau$ | Random effects | Empirical $R^{10}$ | Estimated $R^{10}$ |
|---|---|---|---|
| 0 | None | (3.64,3.69)* | (4.29,4.40)** |
|   | yes |  | (4.28,4.39) |
| 1 | None | (3.64, 3.77) | (4.93,5.30) |
|   | yes |  | (4.23,4.52) |
| 2 | None | (3.62,3.83) | (5.86,6.67) |
|   | yes |  | (4.17,4.70) |
| 3 | None | (3.59,3.91) | (6.84,8.12) |
|   | yes |  | (3.59,4.72) |
| 4 | None | (3.57,4.00) | (7.84,9.57) |
|   | yes |  | (3.62,4.90) |

* 95% confidence interval over 25 simulated samples. **95% confidence interval for the mean of estimate over 25 simulated samples.

# 6 Application

In this section we apply the random effects model to two types of data. First we analyze the maximums of the financial series of daily stock market's return value collected from EuroXX, FTS, HS, Nikkei, SMI, and SP indexes. Second we analyze maximum temperature and precipitation in the city of Abbotsford in the province of British Columbia in Canada.

## 6.1 Analysis of maximum return value of indexes

We analyze daily returns from six stock market's indexes downloaded from www.unige.ch/ses/metri/gilli/evtrm/. We have considered the maximum of the total percent change in a given stock market's value for each year of the six stock indexes. Table 6.1 shows some descriptive statistics for these indexes. The mean and standard deviation of yearly maximum of changes are highest for HS and lowest for SP.

**Table 6.1:** Description of maximum return of indexes.

| Index | Description | Year | Mean* | SD* |
|---|---|---|---|---|
| EuroXX | Dow Jones EuroXX stock | 1987 - 2004 | 4.47 | 1.91 |
| FTSE | FTSE 100 stock | 1984 - 2004 | 3.32 | 1.56 |
| HS | Hang Seng stock | 1981 - 2004 | 6.28 | 3.22 |
| Nikkei | Nikkei 225 stock | 1970 - 2004 | 4.55 | 2.32 |
| SMI | Swiss Market stock | 1988 - 2004 | 4.15 | 1.81 |
| SP | S&P 500 stock | 1960 - 2004 | 3.10 | 1.48 |

*Mean and standard deviation of yearly maximums.

### 6.1.1 Joint modeling of yearly maximums

We fit the GEV distribution to the yearly maximum return values of the combined six indexes available during the period of times mentioned in Table 6.1. We allow the random effects changes between indexes and between years. The results are reported in Table 6.2. The variance of the random effects is estimated positive in both random effects models. This indicates the existence of heterogeneity in maximum return values among indexes and among years. The empirical estimate of $R^{10}$ for yearly maximums is 7.06 and is within the *95% HPD* interval of $R^{10}$ produced by all models. The point estimate of $R^{10}$ from fixed effects model is 7.24 and from index random effects model is 7.03. The estimates from index random effects models are very close to the empirical estimate. *DIC* for fixed random effects, index random effects, and yearly random effects models are 665.845, 643.024, and 609.960 and the *log-likelihoods* are -331.43, -317.91, and -290.14 respectively. According to *DIC* and *log-likelihood* criteria both random effects models work better than fixed effects model and the yearly random effects model fits the data overall better than index random effects model. The yearly random effects model indicates that in one year out of the next 10 years the return value exceeds 6.44% while this limit for index random effects is 7.03%. Since unobservable factors affect indexes differently we believe that the return values are correlated among the indexes and the joint modeling produces more reliable estimates than separate modeling for the return level when an overall interpretation of the stock market's return level is of interest. Separate modeling of indexes is useful when focus is on individual index independently from the other indexes.

**Table 6.2:** Result from applying the block maxima method to combined yearly maximums from six indexes.

| Subject | Parameter | Mean | SD | LB* | UB* |
|---|---|---|---|---|---|
| None | $\xi$ | 0.17 | 0.08 | 0.03 | 0.32 |
|  | $\mu$ | 3.07 | 0.14 | 2.81 | 3.34 |
|  | $R^{10}$ | 7.24 | 0.47 | 6.35 | 8.15 |
|  | $\sigma$ | 1.51 | 0.11 | 1.29 | 1.74 |
| Index | $\xi$ | 0.15 | 0.08 | 0.01 | 0.30 |
|  | $\mu$ | 3.26 | 0.37 | 2.55 | 3.94 |
|  | $R^{10}$ | 7.03 | 0.58 | 6.02 | 8.26 |
|  | $\sigma$ | 1.40 | 0.10 | 1.19 | 1.59 |
|  | $\tau^2$ | 0.85 | 1.32 | 0.04 | 2.28 |
| Year | $\xi$ | 0.27 | 0.10 | 0.09 | 0.47 |
|  | $\mu$ | 3.03 | 0.19 | 2.65 | 3.39 |
|  | $R^{10}$ | 6.44 | 0.49 | 5.51 | 7.37 |
|  | $\sigma$ | 1.10 | 0.12 | 0.88 | 1.34 |
|  | $\tau^2$ | 1.08 | 0.39 | 0.41 | 1.85 |

*LB and LU are Lower and upper 95% HPD interval.

**Figure 6.1:** Quantile and return level plots of fitting yearly random effects (left) and index random effects (right) GEV models to yearly maximum return value for combined six indexes.

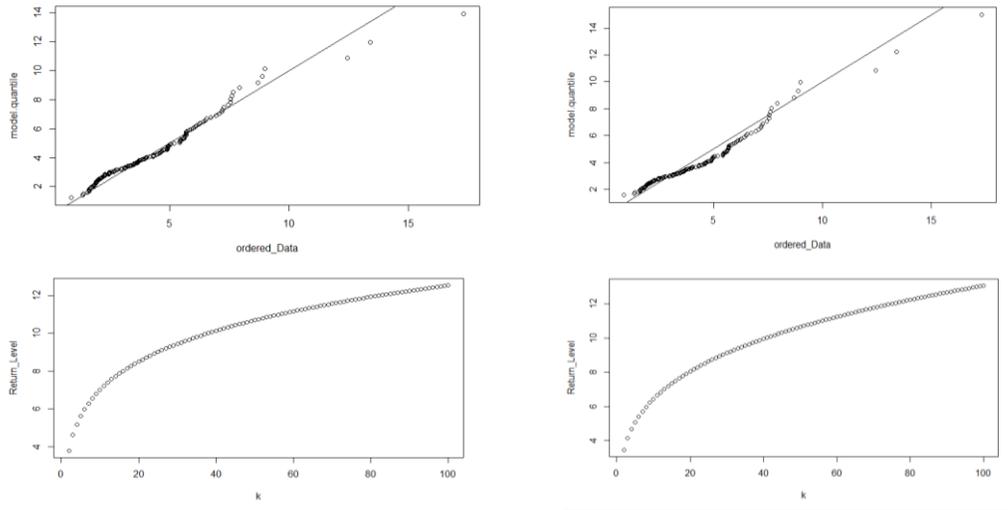

### 6.1.2 Separate modeling of yearly maximum

In this section we fit the GEV distribution to yearly maximum return values of indexes separately. We allow the random effects vary between years. Table 6.3 reports the *DIC* for random effects and fixed effects models across indexes. As it is clear from this table random effects model produces smaller *DIC* for all indexes. This indicates that random effects model fit the yearly maximum return values better than fixed effects model for all indexes. The estimates of the parameters produced by random effects models are shown in Tables 6.4. The variance of the random effects, $\tau^2$, is estimated positive for all indexes. This indicates the existence of heterogeneity of maximum return values between years. Manfred Gilli and Evis KÄellezi [3] have analyzed the yearly return values of SP index by using *MLE* method. They have reported an estimate of 4.981 with a 95% confidence interval (4.230,6.485). Our analysis for the SP index in Table 6.4 shows almost the same result as they reported. The empirical value of $R^{10}$ for EuroXX, FTSE, HS, Nikkei, SMI, and SP are 7.76, 5.81, 11.20, 7.59, 6.68, and 4.93 respectively. Table 6.4 shows that the empirical values are located in the 95% *HPD* interval for all indexes. This indicates that the random effects model performs well to estimate $R^{10}$ for all indexes.

**Table 6.3:** DIC for applying block maxima method to yearly maximums return values with yearly random effects.

| Random effects | EUROXX | FTSE  | HS     | NIKKei | SMI   | SP     |
|----------------|--------|-------|--------|--------|-------|--------|
| Yes            | 69.76  | 69.80 | 96.08  | 147.72 | 29.74 | 152.97 |
| None           | 75.52  | 74.54 | 114.44 | 152.20 | 72.40 | 155.10 |

**Table 6.4:** Result from applying the block maxima method to yearly maximums return values with yearly random effects.

| Index | Parameter | Mean | SD | LB* | UB* |
|---|---|---|---|---|---|
| EuroXX | $\xi$ | -0.66 | 0.50 | -1.56 | 0.43 |
|  | $\mu$ | 3.99 | 0.71 | 2.55 | 5.33 |
|  | $R^{10}$ | 6.45 | 0.71 | 4.97 | 7.92 |
|  | $\sigma$ | 2.34 | 1.10 | 0.53 | 4.53 |
|  | $\tau^2$ | 0.78 | 1.21 | 0.00 | 3.16 |
| FTSE | $\xi$ | 0.44 | 0.38 | -0.21 | 1.23 |
|  | $\mu$ | 2.57 | 0.28 | 2.07 | 3.15 |
|  | $R^{10}$ | 6.10 | 1.60 | 3.39 | 9.30 |
|  | $\sigma$ | 0.95 | 0.32 | 0.36 | 1.61 |
|  | $\tau^2$ | 0.24 | 0.43 | 0.00 | 1.02 |
| HS | $\xi$ | 0.69 | 0.55 | -0.07 | 1.80 |
|  | $\mu$ | 4.87 | 0.39 | 4.14 | 5.67 |
|  | $R^{10}$ | 10.52 | 2.09 | 7.23 | 14.75 |
|  | $\sigma$ | 1.35 | 0.60 | 0.31 | 2.43 |
|  | $\tau^2$ | 0.90 | 0.78 | 0.01 | 2.30 |
| Nikkei | $\xi$ | 0.25 | 0.23 | -0.13 | 0.69 |
|  | $\mu$ | 3.44 | 0.32 | 2.80 | 4.06 |
|  | $R^{10}$ | 8.12 | 1.43 | 5.84 | 11.19 |
|  | $\sigma$ | 1.53 | 0.37 | 0.71 | 2.25 |
|  | $\tau^2$ | 0.35 | 0.63 | 0.00 | 1.70 |
| SMI | $\xi$ | -0.41 | 0.53 | -1.35 | 0.55 |
|  | $\mu$ | 3.88 | 0.62 | 2.70 | 4.75 |
|  | $R^{10}$ | 5.41 | 1.00 | 4.59 | 7.66 |
|  | $\sigma$ | 1.20 | 0.84 | 0.13 | 2.54 |
|  | $\tau^2$ | 2.00 | 1.83 | 0.01 | 5.31 |
| SP | $\xi$ | 0.18 | 0.19 | -0.14 | 0.53 |
|  | $\mu$ | 2.41 | 0.19 | 2.06 | 2.79 |
|  | $R^{10}$ | 5.25 | 0.85 | 3.88 | 6.79 |
|  | $\sigma$ | 1.01 | 0.19 | 0.60 | 1.38 |
|  | $\tau^2$ | 0.14 | 0.20 | 0.00 | 0.55 |

*LB and LU are lower and upper bounds of 95% HPD interval.

## 6.2 Analysis of precipitation

We have considered the data on daily precipitation in the city of Abbotsford in the province of British Columbia in Canada from 1994 to 2015. Data are collected from historical climate data section of the web page https://www.canada.ca/en/services/environment/weather/data-research.html. Plots in Figure 6.2 report daily and yearly maximum precipitation. The maximum and minimum precipitation occurred in 2003 and 2000 respectively. Our analysis indicates that the presence of random effects in scale parameter does not work well for this data set. We therefore continue with location random effects model. Table 6.5 reports the results of fitting fixed effects and random effects models to yearly maximum precipitation. The *log-likelihood* and *DIC* for fixed effects model are -93.34 and 188.98 while for random effects model are -93.31 and 189.01, respectively. These criteria do not show any evidence that the random effects model fits the yearly maximum precipitation better than fixed effects model. The empirical estimate of $R^{10}$ for yearly maximum precipitation is calculated as 83.2 mm, which is located in all 95% HPD interval of the estimate of $R^{10}$. The random effects model indicates that in one year out of the next 10 years the total precipitation in this location exceeds 87.86mm. The quantile plots of fitted models are reported in Figure 6.4 and support the use of any of these two models. These models estimate the parameters almost equally.

**Figure 6.2:** Time series plots of precipitation in Abbotsford.

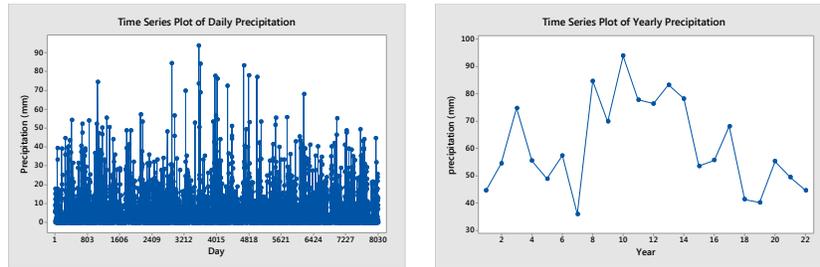

**Table 6.5:** Result from applying the block maxima method to yearly maximum precipitation in Abbotsford.

| Random effects | Parameter | Mean | SD | LB* | UB* |
|---|---|---|---|---|---|
| None | $\xi$ | -0.08 | 0.23 | -0.53 | 0.35 |
| | $\mu$ | 54.08 | 3.93 | 46.50 | 61.88 |
| | $R^{10}$ | 88.12 | 10.40 | 71.89 | 111.40 |
| | $\sigma$ | 16.05 | 3.36 | 10.17 | 22.60 |
| Yes | $\xi$ | -0.09 | 0.24 | -0.53 | 0.41 |
| | $\mu$ | 54.09 | 3.97 | 46.50 | 62.07 |
| | $R^{10}$ | 87.86 | 9.44 | 72.79 | 107.90 |
| | $\sigma$ | 16.31 | 3.61 | 10.07 | 23.87 |
| | $\tau^2$ | 2.22 | 12.02 | 0.00 | 8.94 |

*LB and LU are Lower and upper bounds of 95% HPD interval.

**Figure 6.4:** Quantile plots of fixed effects (left) and random effects (right) models.

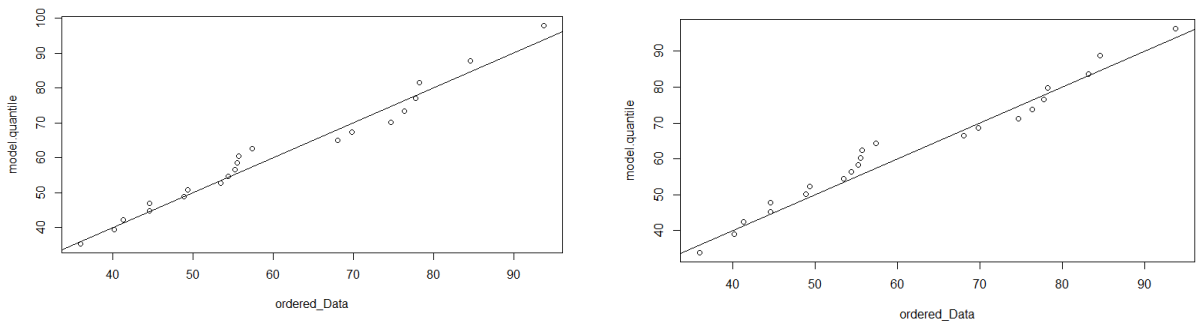

## 6.3 Analysis of temperature

Our next application is the analysis of the yearly minimum temperature in Abbotsford from 1994 to the end of 2015 in Abbotsford in British Columbia. Figure 6.5 reports the time series on daily and yearly minimum temperature. The yearly minimum temperature ranges from -14.8º C to -4.2º C with minimum occurred in 2012. Tale 6.6 reports the result of fitting the GEV distribution to the negative of the yearly minimum temperature. The *log-likelihood* and *DIC* for fixed effects model are -53.84 and 110.03 and for random effects model are -49.93 and 102.95, respectively. Both *log-likelihood* and *DIC* statistics confirm that the GEV random effects model performs better than GEV fixed effects model. The empirical estimate of $R^{10}$ for the negative of the yearly minimum temperature is calculated as 12.80º C. The estimate of $R^{10}$ by the random effects is 12.83 while by fixed effects model is 13.55. This indicates that the random effects model performs better than fixed effects model for estimation of the return level. Our random effects model indicates that in one year out of the next 10 years the temperature goes below -12.83º C in this region. Figure 6.6 shows the quantile and return level plots of applying GEV random effects model to yearly minimum temperature.

**Figure 6.5:** Time series plot of minimum temperature.

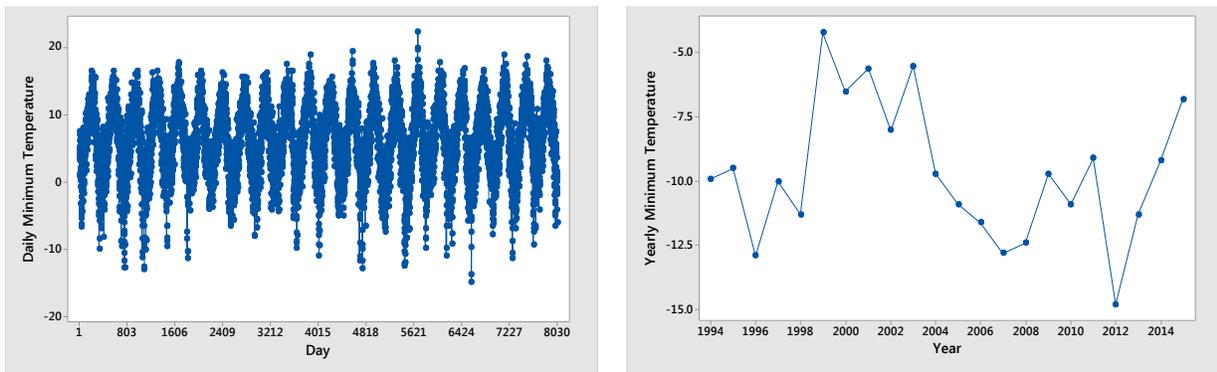

**Table 6.6:** Result from applying the block maxima method to negative of the yearly minimum temperature.

| Random effects | Parameter | Mean | SD | LB* | UB* |
|---|---|---|---|---|---|
| None | $\xi$ | -0.35 | 0.17 | -0.66 | 0.02 |
|  | $\mu$ | 8.70 | 0.72 | 7.25 | 10.02 |
|  | $R^{10}$ | 13.55 | 0.88 | 11.95 | 15.33 |
|  | $\sigma$ | 3.10 | 0.63 | 2.07 | 4.31 |
| Yes | $\xi$ | -0.58 | 0.34 | -1.30 | -0.03 |
|  | $\mu$ | 9.00 | 0.79 | 7.33 | 10.39 |
|  | $R^{10}$ | 12.83 | 0.87 | 11.37 | 14.47 |
|  | $\sigma$ | 2.95 | 0.76 | 1.73 | 4.39 |
|  | $\tau^2$ | 1.02 | 0.10 | 0.82 | 1.23 |

*LB and LU are Lower and upper bounds of 95% HPD interval.

**Figure 6.6:** Quantile (left) and return level (right) plots of random effects GEV model applied to yearly minimum temperature.

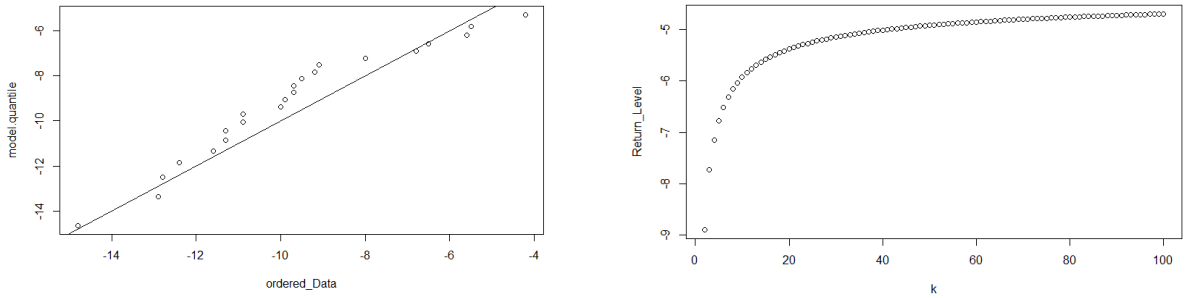

## 7 Informative prior

In many applications data are massive and are continuously collected over time. There is always a possibility of missing part or even the entire of data. One strategy is to use as much data as are available using MLE or Bayesian method. In this case we lose the information included in missed data. One advantage of using Bayesian method using *MCMC procedure* is the possibility of saving the information included in the past data into the posterior distribution of the parameters and uses them in the analysis of current data. We discuss this property in using both BM and POT methods for analyzing extremes.

### 7.1 Block maxima method

### 7.1.1 Simulation

We set up a simulation study based on the estimates of the parameters of the GEV distribution obtained in the analyzing of SP index in Table 6.4. We produced 58 observations (analogues to 58 years SP index data) from GEV distribution with parameters $\xi = 0.18, \mu = 2.41, \sigma = 1.01$. As the GEV distribution is not included in SAS software use the inverse transformation sampling method for generating random numbers from this distribution. We divided the simulated data in two parts. The first part includes 43 observations (analogues to the yearly maximum data we fitted for SP index in section 6) and the second part includes 15 observations. We used uninformative prior distributions using *MCMC procedure* for the first part of the data and saved the posterior distribution of the parameters. We then fitted the GEV distribution to the second part of the simulated data using uninformative prior and informative prior distributions. The posterior distributions of the parameters obtained from the first part of data are shown in Figure 7.1. The diagnostics plots show that the posterior distributions of $b$ ($\xi$) and $mu$ ($\mu$) are close to a normal distribution and the posterior distribution of sigma ($\sigma$) is close to a gamma distribution. So we have considered informative priors distributions $\xi$ ~ *normal(mean=0.08, sd=0.05)*, $\mu$ ~ *normal(mean=2.58,sd=0.06)*, and $\sigma$ ~ *gamma(shape=35.79, scale=0.03)*. The simulation results are reported in Table 7.1. As the return level in the analysis of the yearly return values is often of interest we have calculated the true and estimated values of $R^k$ and reported in Figure 7.2. It is clear from this plot that the uninformative prior distribution model always overestimates the return level and the informative prior distribution approach produces less biased estimate.

**Figure 7.1:** The *MCMC procedure* result for 43 simulated observations. Note that $\xi$ is shown as b and $\mu$ as *mu*.

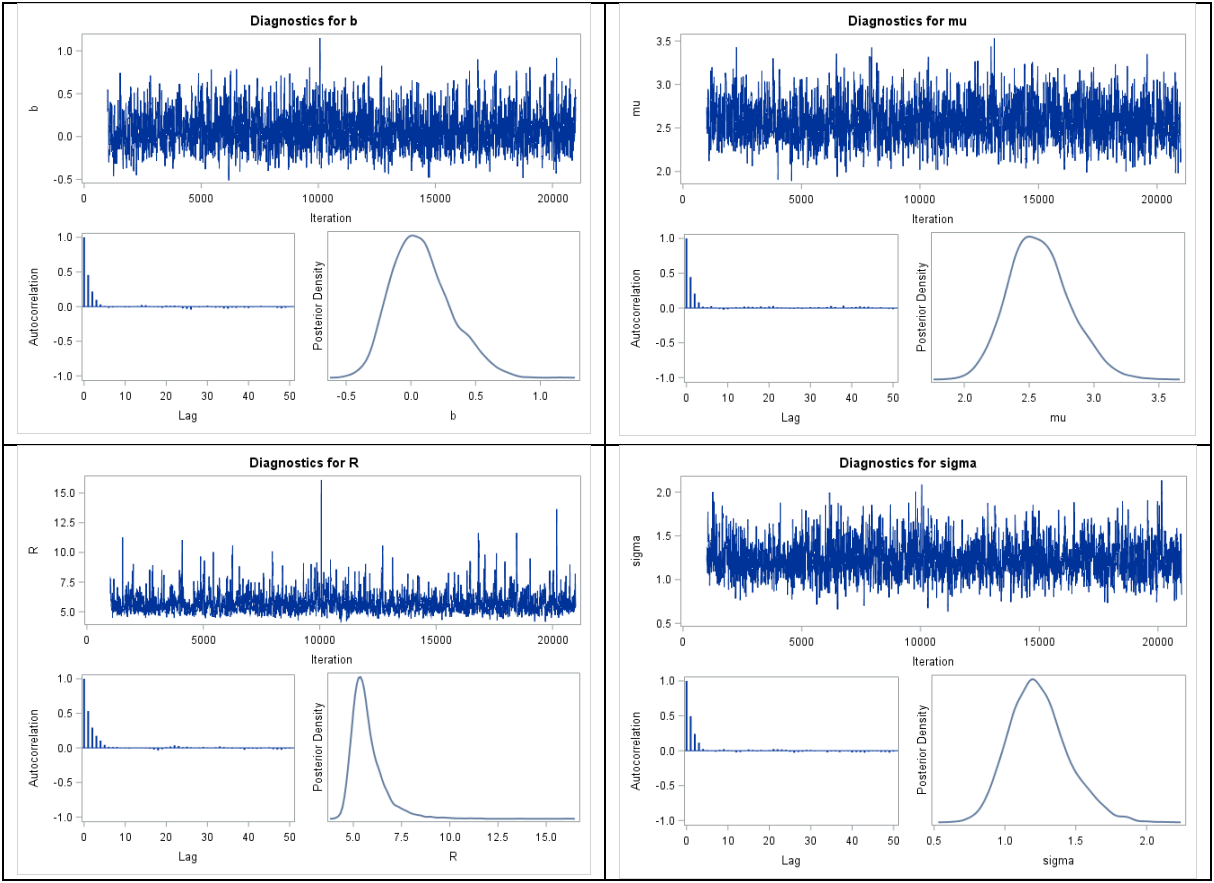

**Table 7.1:** Simulation result for true values $\xi = 0.18, \mu = 2.41, \sigma = 1.01$.

| Data | Parameter | Mean | SD | LB* | UB* |
|---|---|---|---|---|---|
| Years 1-43 | $\xi$ | 0.08 | 0.23 | -0.33 | 0.55 |
| Uninformative prior | $\mu$ | 2.58 | 0.24 | 2.12 | 3.03 |
|  | $R^{10}$ | 5.73 | 0.89 | 4.48 | 7.49 |
|  | $\sigma$ | 1.24 | 0.21 | 0.86 | 1.66 |
| Years 44-58 | $\xi$ | 0.05 | 0.19 | -0.33 | 0.43 |
| Informative prior | $\mu$ | 2.67 | 0.20 | 2.29 | 3.05 |
|  | $R^{10}$ | 5.48 | 0.74 | 4.29 | 6.97 |
|  | $\sigma$ | 1.15 | 0.17 | 0.84 | 1.49 |
| Years 44-58 | $\xi$ | 0.01 | 0.45 | -0.82 | 0.89 |
| Uninformative prior | $\mu$ | 2.83 | 0.38 | 2.16 | 3.59 |
|  | $R^{10}$ | 6.19 | 2.96 | 4.05 | 10.07 |
|  | $\sigma$ | 1.28 | 0.40 | 0.58 | 2.09 |
| Years 44-58 | $\xi$ | 0.00 | 0.43 | -0.92 | 0.92 |
| MLE method | $\mu$ | 2.80 | 0.35 | 2.05 | 3.55 |
|  | $R^{10}$ | 5.02 | 0.72 | 3.48 | 6.55 |
|  | $\sigma$ | 0.99 | 0.29 | 0.37 | 1.60 |

*LB and LU are Lower and upper bounds of 95% HPD interval.

**Figure 7.2:** Comparison of models in estimation of the return levels for years 44 to 58.

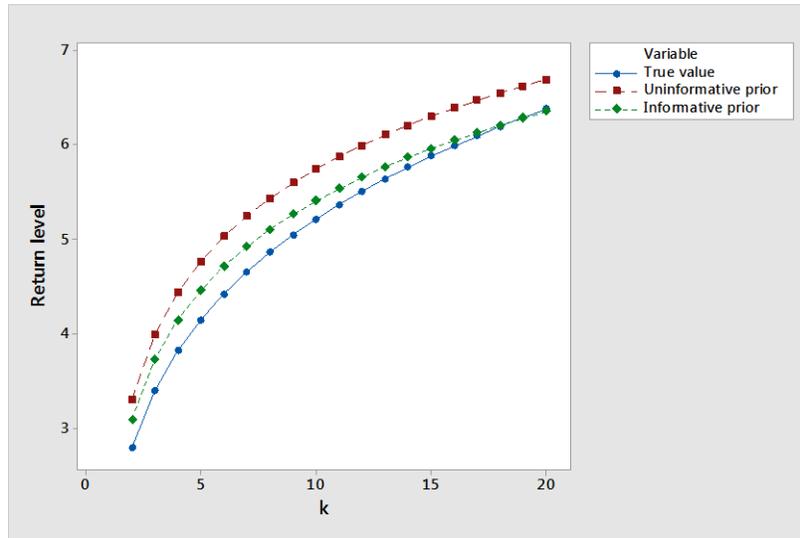

In previous simulation we simulated random data directly from GEV distribution and investigated on the importance of using informative prior distribution. We set up another simulation study in which random effects affect the original observations within blocks. We investigate the importance of random effects modeling simultaneously with using the informative prior distribution. We have produced daily observations for 12 periods of 5 years each. The observations produced from a normal distribution with mean $0.02 + \delta_i$ and standard deviation 1.24 (0.02 and 1.24 are obtained from six indexes that we used in section 6). We assume the random variable $\delta_i$ has a normal distribution with mean zero and standard deviation 1 varying between periods. We split the simulated data in two parts. The first part includes the first 45 years (9 periods) and the second part includes the next 15 years (periods 10, 11, and 12). We have fitted the GEV model to both datasets with and without random effects using informative and uninformative priors. The informative prior distributions used in fitting the 15 years data by random effects (fixed effects) GEV model are collected from the posterior distributions of applying the random effects (fixed effects) GEV model to the first 45 years data. The results are reported in Tables 7.2. Regardless of using informative or uninformative priors the random effects model produces larger *log-likelihood* and smaller *DIC* than fixed effects model, which indicates the advantage of using random effects model. The informative priors obtained from the first 45 years are not effective to improve the *log-likelihood* and *DIC* in fitting the next 15 years with or without applying random effects. The average of the empirical estimate of $R^{10}$ calculated over 3 periods 10, 11, and 12 is 4.26 which is only located in the 95% confidence interval of $R^{10}$ estimated by informative random effects model.

**Table 7.2:** Result from applying the block maxima method to 12 periods of 5 years each simulated data. Figures shown are the average over the 25 replications.

| Data and prior | Parameter | Mean | SE | LB* | UB* |
|---|---|---|---|---|---|
| First 45 years data | $\xi$ | -0.29 | 0.25 | -0.38 | -0.19 |
| Uninformative prior | $\mu$ | 3.38 | 0.4 | 3.22 | 3.54 |
| without random effects | $R^{10}$ | 5.10 | 0.5 | 4.89 | 5.31 |
| | $\sigma$ | 1.03 | 0.3 | 0.90 | 1.17 |
| | $ll^{**}$ | -61.75 | 10.6 | -66.13 | -57.37 |
| | DIC | 126.05 | 21.15 | 117.31 | 134.80 |
| First 45 years data | $\xi$ | -0.10 | 0.15 | -0.16 | -0.04 |
| Uninformative prior | $\mu$ | 3.52 | 0.35 | 3.37 | 3.67 |
| with random effects | $R^{10}$ | 4.40 | 0.4 | 4.23 | 4.58 |
| | $\sigma$ | 0.43 | 0.05 | 0.41 | 0.46 |
| | $\tau^2$ | 1.25 | 0.85 | 0.90 | 1.59 |
| | $ll$ | -28.34 | 6.35 | -30.96 | -25.73 |
| | DIC | 65.84 | 12.7 | 60.60 | 71.08 |
| Last 15 years data | $\xi$ | -0.30 | 0.3 | -0.42 | -0.19 |
| Uninformative prior | $\mu$ | 3.29 | 0.5 | 3.08 | 3.50 |
| without random effects | $R^{10}$ | 5.15 | 0.95 | 4.76 | 5.54 |
| | $\sigma$ | 1.09 | 0.5 | 0.88 | 1.30 |
| | $ll$ | -18.95 | 6.15 | -21.49 | -16.40 |
| | DIC | 39.71 | 12.15 | 34.69 | 44.73 |
| Last 15 years data | $\xi$ | -0.23 | 0.7 | -0.52 | 0.05 |
| Uninformative prior | $\mu$ | 3.58 | 0.6 | 3.33 | 3.83 |
| with random effects | $R^{10}$ | 5.05 | 1.5 | 4.44 | 5.66 |
| | $\sigma$ | 0.69 | 0.85 | 0.35 | 1.03 |
| | $\tau^2$ | 5.61 | 8.75 | 1.99 | 9.22 |
| | $ll$ | -9.61 | 3.45 | -11.05 | -8.18 |
| | DIC | 22.53 | 6.25 | 19.83 | 25.23 |
| Last 15 years data | $\xi$ | -0.28 | 0.2 | -0.36 | -0.20 |
| Informative prior | $\mu$ | 3.33 | 0.4 | 3.16 | 3.50 |
| without random effects | $R^{10}$ | 4.95 | 0.75 | 4.63 | 5.27 |
| | $\sigma$ | 0.94 | 0.35 | 0.80 | 1.08 |
| | $ll$ | -18.64 | 6.05 | -21.12 | -16.15 |
| | DIC | 39.05 | 12 | 34.09 | 44.01 |
| Last 15 years data | $\xi$ | -0.10 | 0.2 | -0.19 | -0.02 |
| Informative prior | $\mu$ | 3.45 | 0.25 | 3.34 | 3.55 |
| with random effects | $R^{10}$ | 4.37 | 0.35 | 4.22 | 4.52 |
| | $\sigma$ | 0.44 | 0.1 | 0.41 | 0.47 |
| | $\tau^2$ | 2.60 | 2.45 | 1.58 | 3.62 |
| | $ll$ | -9.53 | 2.85 | -10.70 | -8.36 |
| | DIC | 22.44 | 5.65 | 20.06 | 24.82 |

*LB and LU are Lower and upper bounds of 95% confidence interval. **$ll$ is the log-liklihood

### 7.1.2 Application

We have used daily return values of SP index from 1960 to 2018. We use the posterior distributions $\xi \sim normal(mean=0.18, sd=0.19)$, $\mu \sim normal(mean=2.41, sd=0.19)$, and $\sigma \sim gamma(shape=28.29, scale=0.04)$ obtained from the analysis of data from 1960 to 2004 reported in table 6.4 as informative prior distributions to model the data from August 2004 to November 2018. We let the random effects vary between years. The results of applying informative and uninformative prior distributions are reported in Table 7.3. The *log-likelihood* and *DIC* do not distinguish between two models. The empirical estimate of $R^{10}$ is 61.9 and is closer to the estimate from informative prior model. Combining this analysis with the result of the simulation result recommends the use of informative prior random effects model.

As another application we have considered daily precipitation in Abbotsford from 1994 to 2018. We have analyzed this data from 1994 to 2015 in section 6. We have divided the data into two parts. The first part is from 1994 to 2008 and the second part from 2009 to 2018. As figure 6.3 shows a seven-year cyclical variation in yearly precipitation, we allowed the random effects vary every seven years in this analysis. We have fitted the GEV random effects model to the data from 1994 to 2008. We have then used the posterior distributions distributions $\xi \sim normal(mean=-0.42, sd=0.52)$, $\mu \sim normal(mean=61.66, sd=5.34)$, and $\sigma \sim gamma(shape=3.08, scale=3.41)$ obtained from this fit as prior distributions for fitting the GEV random effects model to the rest of data from 2009 to 2018. The results are reported in Table 7.4. The empirical value of $R^{10}$ for the data from 2009 to 2018 is 61.9mm. The estimate from informative random effects model is 70.31 and from uninformative random effects model is 85.90. This indicates that using informative priors produce better estimate for the return level.

**Table 7.3:** Result from applying the block maxima method to yearly maximum return value for SP index from 2004 to 2018.

| Prior | Parameter | Mean | SD | LB* | UB* |
|---|---|---|---|---|---|
| Uninformative | $\xi$ | 0.79 | 0.59 | -0.06 | 2.04 |
|  | $\mu$ | 2.35 | 0.38 | 1.61 | 3.14 |
|  | $R^{10}$ | 5.54 | 0.60 | 4.60 | 6.92 |
|  | $\sigma$ | 1.11 | 0.63 | 0.16 | 2.34 |
|  | $\tau^2$ | 0.49 | 0.67 | 0.00 | 1.48 |
| Informative | $\xi$ | 0.27 | 0.13 | 0.02 | 0.54 |
|  | $\mu$ | 2.38 | 0.16 | 2.06 | 2.70 |
|  | $R^{10}$ | 5.67 | 0.74 | 4.40 | 7.12 |
|  | $\sigma$ | 1.05 | 0.16 | 0.74 | 1.38 |
|  | $\tau^2$ | 0.34 | 0.54 | 0.00 | 1.12 |

*LB and LU are Lower and upper bounds of 95% HPD interval.

**Table 7.4:** Result from applying the block maxima method to yearly maximum precipitation in Abbotsford from 2009 to 2018.

| Prior | Parameter | Mean | SD | LB* | UB* |
|---|---|---|---|---|---|
| Uninformative | $\xi$ | 0.19 | 0.56 | -0.80 | 1.41 |
|  | $\mu$ | 46.11 | 4.66 | 37.37 | 56.20 |
|  | $R^{10}$ | 85.90 | 75.98 | 51.63 | 180.00 |
|  | $\sigma$ | 9.41 | 4.25 | 3.51 | 17.51 |
|  | $\tau^2$ | 106.20 | 1755.40 | 0.00 | 163.30 |
| Informative | $\xi$ | -0.17 | 0.31 | -0.78 | 0.43 |
|  | $\mu$ | 50.85 | 3.15 | 44.71 | 57.03 |
|  | $R^{10}$ | 70.31 | 10.93 | 57.66 | 89.24 |
|  | $\sigma$ | 9.74 | 2.66 | 5.41 | 15.25 |
|  | $\tau^2$ | 1.00 | 0.10 | 0.80 | 1.19 |

*LB and LU are Lower and upper bounds of 95% HPD interval.

## 7.2 Peak over threshold method

In this section we investigate the importance of using the informative priors in the analysis of extremes using POT method. We report some simulation studies and application to real data. As the location of GPD is fixed in this method we do not consider random effects.

### 7.2.1 Simulation

We have simulated 25 samples each includes 65 years daily observation from normal distribution with mean 2 and standard deviation 1. As noted in section 2, there is not any proved formula for the selection of threshold. We have assumed the threshold to be $u=4$ in this simulation. With this choice of $u$, roughly 2.5% of the observations in each year would exceed 4. We have divided 65 years simulated data into two parts of 60 years and 5 years. We have applied the POT approach to the first 60 years simulated data using uninformative prior distributions $\xi \sim$ *uniform(-10000, 10000)* and $\sigma \sim$ *gamma (shape=0.0001, scale=10000)* and obtained the following posterior distributions $\xi \sim$ *normal(mean=-0.1294, sd=0.02858)* and $\sigma \sim$ *gamma (shape=989, scale=0.0005)*. We have used these posterior distributions as prior distributions for applying the POT approach to the next 5 years simulated data. We have also used uninformative prior distribution for analyzing the last 5 years simulated data. In order to investigate the effect of informative prior distributions we have analyzed the data in a shorter period of time of only one year (year 61). The estimates of the parameters and the estimates of $VAR_{0.05}$ and $ES_{0.05}$ are reported in Table 7.5. The empirical estimates of $VAR_{0.05}$ and $ES_{0.05}$ are calculated and reported in Table 7.6 for comparison. The informative prior and uninformative prior approaches produce almost equal estimates for $VAR_{0.05}$ and $ES_{0.05}$ for last 5 years data (years 61-65). These estimates are very close to the empirical estimates reported in Table 7.6. The informative prior approach estimates $VAR_{0.05}$ and $ES_{0.05}$ very close to the empirical estimates for the data in year 61 while the uninformative approach estimates are very biased. This simulation indicates that the informative priors obtained from the past data help to produce accurate estimates when data are observed in a short period of time.

### 7.2.2 Application

We have considered the data on daily return values of SP index from 5 January 1960 to 19 November 2018. We fitted the data from 5 January 1960 to 16 August 2004 using uninformative prior distributions. We obtained the following posterior distributions $\xi \sim normal(0.15, sd=0.05)$ and $\sigma \sim gamma\ (shape=255.65, scale=0.0023)$. We used these posterior distributions as prior distributions and fitted the data from 17 August 2004 to 19 November 2018. In order to compare our result with the result from Gilli and Kaellezi [3] we have assumed $u = 1.4$ and $p=0.01$. The results are reported in Table 7.7. Our estimates of $VAR_{0.01}$ and $ES_{0.01}$ for the data from 5 January 1960 to 17 August 2004 are very close to the estimate reported by Gilli and Kaellezi [3]. Their estimates for $VAR_{0.01}$ and $ES_{0.01}$ are 2.505 with 95% confidence interval (2.411, 2.609) and 3.351 with 95% confidence interval (3.151,3.639) respectively. We found *DICs* approximately equal for comparable data sets and so this measure cannot distinguish between models. The empirical estimates of $VAR_{0.01}$ and $ES_{0.01}$ for data from 17 August 2004 to 19 November 2018 are 3.19 and 4.56 respectively. Table 7.7 shows that the informative prior model estimates $ES_{0.01}$ better than the uninformative prior model while the uninformative prior model estimates $VAR_{0.01}$ with less bias than informative prior model. The empirical estimates of $VAR_{0.01}$ and $ES_{0.01}$ for the data from 17 August 2004 to 17 August 2005 are 1.59 and 1.72 respectively. Table 7.7 shows more efficient estimates for these parameters by using informative prior distributions than by using uninformative prior distributions. The diagnostic plots from using the informative and uninformative prior distributions are reported in Figures 7.3 and 7.4. These plots show that the estimates from informative prior distribution are more reliable than the estimates from uninformative prior distributions. Considering the mean, the standard deviation of the estimates, and the diagnostic plots indicates that informative prior model obtained from the past data fits the future data more efficient than uninformative prior model. This interpretation is consistent with the interpretation of the simulation study.

The second application is the analysis of daily precipitation in Abbotsford. We assumed several values for the threshold $u$ and report only for $u = 39mm$. We have used the POT approach using uninformative prior distributions and fitted the daily precipitation from 1994 to 2015. We have used the obtained posterior distributions $\xi \sim normal(0.11, sd=0.16)$ and $\sigma \sim gamma\ (shape=11.61, scale=2.15)$ as informative prior distributions for fitting the daily precipitation in the next three years (2016-2018) and to a shorter period of one year (2016). The results of applying informative and uninformative prior distributions are reported in table 7.8. The empirical values of $VAR_{0.05}$ and $ES_{0.05}$ for the data collected in 3 years 2016-2018 are 21.30 and 32.10 and for year 2016 are 18.90 and 31.93 respectively. The uninformative prior models estimate of $VAR_{0.05}$ either very biased or even negative. The informative prior models estimates this parameter better than uninformative models. The uninformative prior models estimate $ES_{0.05}$ very biased but the informative prior models estimate this parameter much better and very close to the empirical values. This application shows that the informative prior model works better than uninformative prior model especially for the data collected in a short period of time.

**Table 7.5:** Simulation result for POT approach. Figures shown are the mean of estimates over 25 replications. Threshold $u$ and probability $p$ are assumed to be 4 and 0.05.

| Data | Parameter | Mean | SD | LB* | LU* |
|---|---|---|---|---|---|
| years 1-60 | $\xi$ | -0.10 | 0.05 | -0.12 | -0.08 |
| | $\sigma$ | 0.41 | 0.02 | 0.40 | 0.42 |
| | $VAR_p$ | 3.66 | 0.03 | 3.64 | 3.67 |
| | $ES_p$ | 4.07 | 0.02 | 4.06 | 4.07 |
| Years 61-65 and informative prior | $\xi$ | -0.13 | 0.00 | -0.14 | -0.13 |
| | $\sigma$ | 0.49 | 0.00 | 0.49 | 0.49 |
| | $VAR_p$ | 3.60 | 0.08 | 3.57 | 3.63 |
| | $ES_p$ | 4.08 | 0.07 | 4.05 | 4.11 |
| Year 61-65 and uninformative prior | $\xi$ | -0.09 | 0.18 | -0.16 | -0.01 |
| | $\sigma$ | 0.45 | 0.08 | 0.42 | 0.48 |
| | $VAR_p$ | 3.63 | 0.15 | 3.57 | 3.69 |
| | $ES_p$ | 4.08 | 0.08 | 4.05 | 4.11 |
| Year 61 and informative prior | $\xi$ | 0.10 | 0.01 | -0.11 | -0.10 |
| | $\sigma$ | 0.41 | 0.00 | 0.41 | 0.42 |
| | $VAR_p$ | 3.63 | 0.23 | 3.53 | 3.72 |
| | $ES_p$ | 4.04 | 0.21 | 3.95 | 4.12 |
| Year 61 and uninformative prior | $\xi$ | 0.05 | 1.27 | -0.47 | 0.57 |
| | $\sigma$ | 0.65 | 0.36 | 0.50 | 0.80 |
| | $VAR_p$ | -0.44 | 19.15 | -8.34 | 7.47 |
| | $ES_p$ | 3.41 | 3.51 | 1.96 | 4.86 |

**Table 7.6:** Empirical estimation of value at risk and expected shortfall for last 15 years simulated data with low heterogeneity. Threshold $u$ and probability $p$ are assumed to be 4 and 0.05.

| Data | Statistics | Mean | SD | LB* | UB* |
|---|---|---|---|---|---|
| Years 61-65 | $VAR_p$ | 3.65 | 0.05 | 3.63 | 3.67 |
| | $ES_p$ | 4.07 | 0.05 | 4.05 | 4.09 |
| Year 61 | $VAR_p$ | 3.65 | 0.14 | 3.59 | 3.71 |
| | $ES_p$ | 4.06 | 0.19 | 3.98 | 4.14 |

**Table 7.7:** Result of applying POT approach to SP daily return values. Threshold and probability $p$ are assumed to be 1.4 and 0.01.

| Data | Parameter | Mean | SD | LB$^*$ | LU$^*$ |
|---|---|---|---|---|---|
| years 1960-2004 | $\xi$ | 0.15 | 0.05 | 0.05 | 0.24 |
| | $\sigma$ | 0.58 | 0.04 | 0.50 | 0.65 |
| | $VAR_p$ | 2.51 | 0.05 | 2.41 | 2.61 |
| | $ES_p$ | 3.38 | 0.13 | 3.16 | 3.63 |
| years 2004-2018 and informative prior | $\xi$ | 0.22 | 0.04 | 0.15 | 0.29 |
| | $\sigma$ | 0.63 | 0.03 | 0.57 | 0.69 |
| | $VAR_p$ | 2.80 | 0.07 | 2.66 | 2.94 |
| | $ES_p$ | 4.01 | 0.20 | 3.62 | 4.40 |
| Years 2004-2018 and uninformative prior | $\xi$ | 0.33 | 0.10 | 0.14 | 0.52 |
| | $\sigma$ | 0.66 | 0.08 | 0.52 | 0.81 |
| | $VAR_p$ | 3.01 | 0.15 | 2.73 | 3.29 |
| | $ES_p$ | 4.88 | 0.70 | 3.82 | 6.12 |
| Year 2004-2005 and informative prior | $\xi$ | 0.14 | 0.05 | 0.05 | 0.24 |
| | $\sigma$ | 0.58 | 0.04 | 0.51 | 0.65 |
| | $VAR_p$ | 1.81 | 0.03 | 1.76 | 1.87 |
| | $ES_p$ | 2.57 | 0.11 | 2.36 | 2.77 |
| Year 2004-2005 and uninformative prior | $\xi$ | 0.69 | 1.76 | -2.28 | 5.17 |
| | $\sigma$ | 0.38 | 0.44 | 0.00 | 1.14 |
| | $VAR_p$ | 1.77 | 1.13 | 1.40 | 2.26 |
| | $ES_p$ | 3.18 | 79.29 | -2.53 | 5.71 |

**Figure 7.3:** MCMC diagnostic plots for the case that informative priors obtained from fitting the data on SP daily return values from 5 January 1960 to 17 August 2004 are used to fit the data from 17 August 2004 to 17 August 2005.

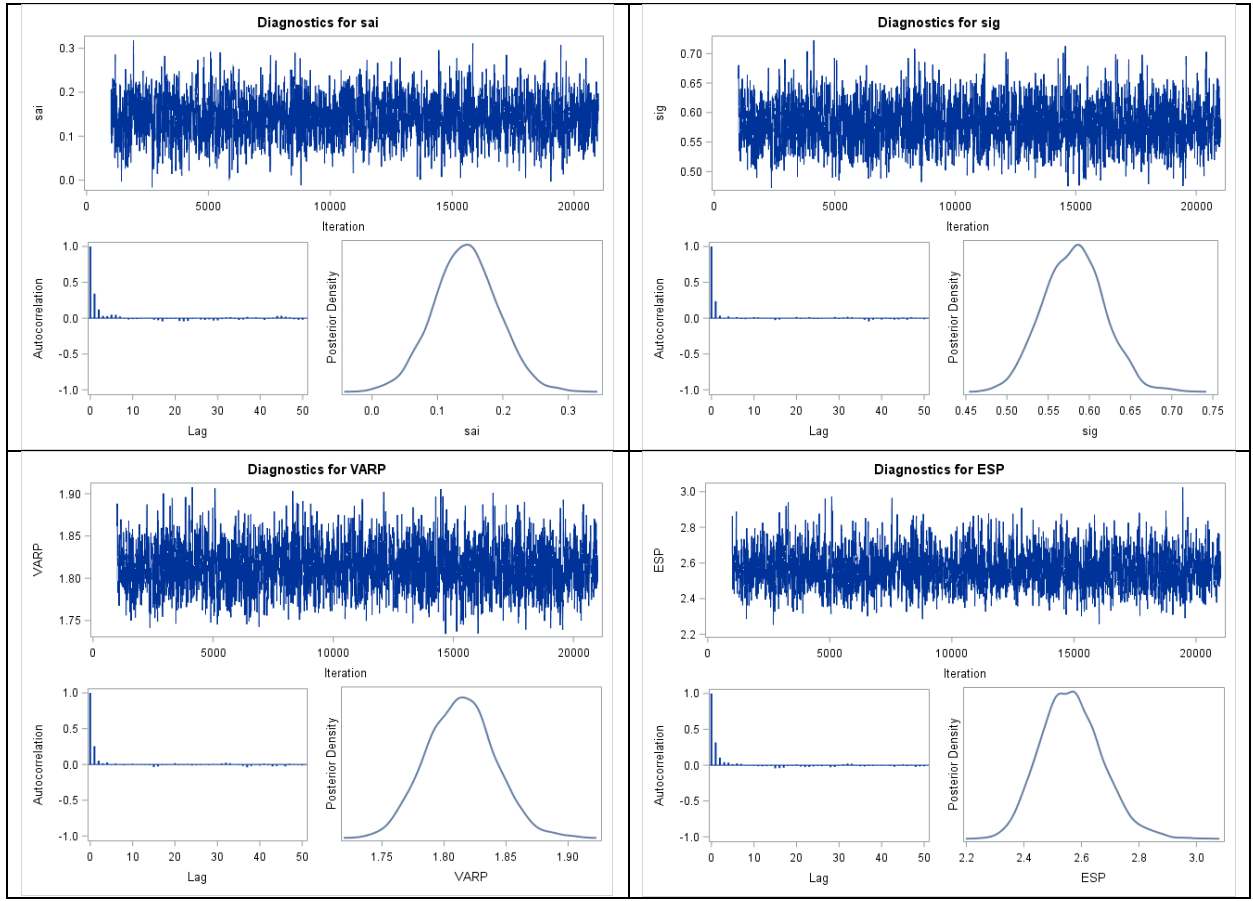

**Figure 7.4:** MCMC diagnostic plots for the case that uninformative priors are used to fit the SP daily returns from 17 August 2004 to 17 August 2005.

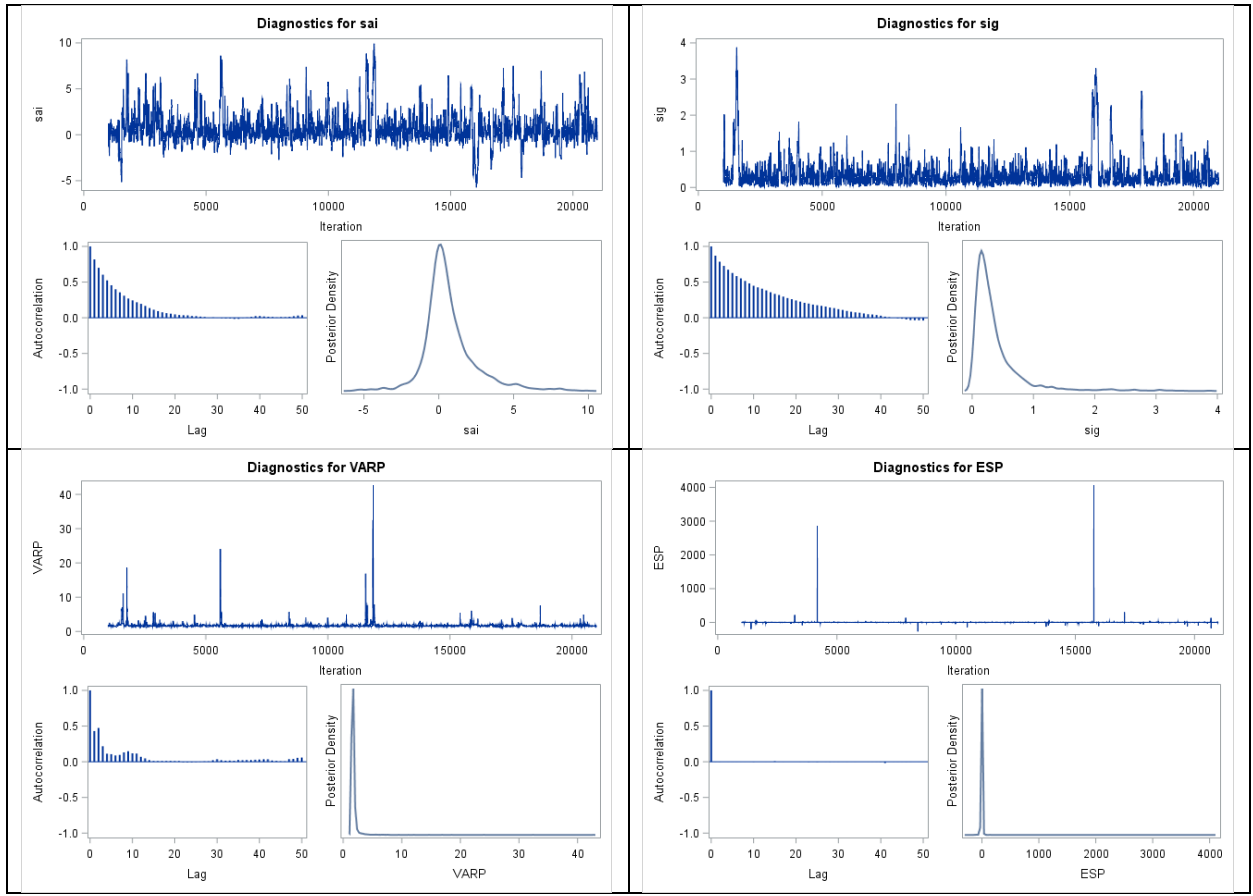

**Table 7.8:** Result of applying POT approach to daily precipitation in Abbotsford. Threshold $u$ and probability $p$ are assumed to be 39 and 0.05.

| Data | Parameter | Mean | SD | LB* | LU* |
|---|---|---|---|---|---|
| years 1994-2015 | $\xi$ | 0.11 | 0.16 | -0.20 | 0.42 |
|  | $\sigma$ | 11.61 | 2.15 | 7.59 | 15.92 |
|  | $VAR_p$ | 22.11 | 4.80 | 12.51 | 30.87 |
|  | $ES_p$ | 33.42 | 2.40 | 28.72 | 37.38 |
| years 2016-2018 and informative prior | $\xi$ | 0.17 | 0.14 | -0.10 | 0.45 |
|  | $\sigma$ | 3.67 | 0.85 | 2.10 | 5.37 |
|  | $VAR_p$ | 34.10 | 1.45 | 31.06 | 36.53 |
|  | $ES_p$ | 37.60 | 0.60 | 36.39 | 38.68 |
| Years 2016-2018 and uninformative prior | $\xi$ | -0.48 | 0.49 | -1.38 | 0.43 |
|  | $\sigma$ | 12.18 | 5.12 | 3.77 | 21.89 |
|  | $VAR_p$ | 4.72 | 45.36 | -53.29 | 36.08 |
|  | $ES_p$ | 27.11 | 13.73 | 5.19 | 38.28 |
| Year 2016 and informative prior | $\xi$ | 0.24 | 0.14 | -0.01 | 0.51 |
|  | $\sigma$ | 2.53 | 0.79 | 1.13 | 4.12 |
|  | $VAR_p$ | 36.16 | 1.02 | 34.29 | 38.03 |
|  | $ES_p$ | 38.65 | 0.33 | 37.97 | 39.26 |
| Year 2016 and uninformative prior | $\xi$ | -0.08 | 1.65 | -3.58 | 3.42 |
|  | $\sigma$ | 16.64 | 13.78 | 0.07 | 45.13 |
|  | $VAR_p$ | -91.89 | 741.70 | -372.50 | 38.98 |
|  | $ES_p$ | 5.73 | 324.20 | -85.09 | 75.46 |

## 8. Summary

We discussed the importance of modeling random effects in the analysis of extreme values using *block maxima* method through simulation and application to return value of economic indexes and climate data. We used Bayesian method and *MCMC procedure* from SAS software for estimation. We used *log-likelihood* and *DIC* as model selection criteria and focused on the estimation of the return level $R^k$ that is a common measure for the prediction of extremes. We compared the estimated value of $R^k$ with its empirical estimate obtained from the sample enumeration. We performed several simulations in which the original observations are produced according to a normal distribution with mean and standard deviation calculated from the real data. We allowed a random variable affects the mean of the normal distribution and vary between periods of time to produce heterogeneous observations. We also replicated each simulation for 25 times and reported the mean of the estimates over the replications to eliminate the possible sampling error. We showed the random effects model fits the simulated data better and estimates $R^{10}$ closer to its empirical value than fixed effects model. We showed that as the standard deviation of the random effects increases the return level is estimated better by the random effects model. For investigation if the simulation results are relevant in practice and if empirical analysis reveals the same patterns and therefore we become reasonably confident we applied the random effects model to analyze the maximums of the financial series of daily stock market's return value collected from EuroXX, FTS, HS, Nikkei, SMI, and SP indexes and daily precipitation and temperature in the city of Abbotsford, BC, Canada. We analyzed the maximum return value through the joint and separate models. The joint modeling of indexes incorporate the correlation between indexes and is useful when an overall estimate of the return level is of interest. The separate modeling of indexed is useful when estimation of the return level for individual index is of interest independent from the other indexes. For joint modeling of indexes we considered two cases. We allowed the random effects vary between indexes in order to capture the heterogeneity among the indexes. We also allowed the random effects changes between blocks (years). Our results show that both random effects models fit the data better than the fixed effects model and the index random effects model produces the best estimate for $R^{10}$. The model selection criteria indicate that random effects model fit the maximum return value of all indexes better than fixed effects model. Our analysis of yearly maximum precipitation does not show any advantage or disadvantage of using random effects model as compare to the fixed effects model but random effects model fits yearly minimum temperature better than the fixed effects model. Our analysis shows that the simulation results and the results from the application to different types of data reveal the same patterns and suggests the use of random effects model in the analysis of extremes by block maxima method. Even when the data are homogeneous the random effects model performs as good as fixed effects model.

We investigated on the effectiveness of using informative prior distributions obtained from the past data on fitting the current data as the second objective of this research. We simulated 58 years data from GEV distribution based on the parameters estimated for SP index data. We obtained the posterior distributions of the parameters using block maxima method applied to the first 43 years data. We then used the obtained posterior distributions as informative prior distributions for the next 15 years data. The uninformative prior model always overestimates the return levels while the informative prior approach produces less biased estimate. We extended the simulation to produce the original data from a normal distribution with random components included in the location parameter for 12 periods each includes 5 years. We obtained the posterior distributions from the first 9 periods with and without considering random effects. We then used these posterior distributions to fit the next 3 periods with and without random effects. We replicated this procedure for 25 times and reported the mean of estimates over the replications. Our analysis shows that the informative priors obtained from the first 9 periods are not effective to improve the *log-likelihood* and *DIC* in fitting the next 3 periods with or without applying random effects. But the average of the empirical estimate of $R^{10}$ calculated over 3 periods 10, 11, and 12 is located only in the 95% confidence interval of $R^{10}$ estimated by informative random effects model. To illustrate the application we have used daily return values of SP index from 1960 to 2018. We used the posterior distributions obtained from data 1960-2004 as informative prior distributions to model the data from August 2004 to November 2018. The model selection criteria do not distinguish between informative prior random effects and uninformative prior random effects models but the return level is estimated with less bias by informative prior random effects model. As another application we have considered daily precipitation in Abbotsford from 1994 to 2018. We have divided the data into two parts from 1994 to 2008 and from 2009 to 2018. Similar analysis indicates that informative prior random effects model produces better estimate for the return level. Simulation results and application to real data confirm that using informative prior together with random effects estimates the return level closer to its empirical value.

Our simulation results show that using the informative prior distributions obtained from the past data makes a considerable difference when *Peak-Over Threshold* method is used in analyzing extremes in a short period of time. The estimates of $VAR_p$ and $ES_p$ from informative and uninformative prior distributions are almost equal for large datasets. The estimates are closer to their empirical values for small datasets when informative prior distributions are used. Application of informative prior distributions obtained from past SP index and Abbotsford precipitation datasets results more reliable estimate for $VAR_p$ and $ES_p$.

In summary, our simulation study and application to real datasets indicate that random effects modeling fits the data better than fixed effects model if heterogeneity of extremes between periods exists. The random effects modeling approach is a safe approach since if the extremes are homogeneous among the periods the results from the random effects model is almost the same as the fixed effects modeling approach. As we do not know about the existence of heterogeneity in real data we recommend using random effects model in block maxima method. Our results show that using informative prior distributions if do not improve the model selection criteria considerably but improve the estimation of the criteria $R^k$, $VAR_p$ and $ES_p$.